\documentclass[aps,showpacs,prd,twocolumn]{revtex4}
\usepackage{amssymb}
\usepackage{amsfonts}
\usepackage{amsmath}
\usepackage{graphicx}
\usepackage{color}
\usepackage[dvips]{epsfig}
\usepackage[dvips]{graphicx}
\usepackage{float}

\begin{document}

\title{Maxwell-Chern-Simons vortices in a CPT-odd Lorentz-violating Higgs
Electrodynamics}
\author{R. Casana, M. M. Ferreira Jr., E. da Hora, A. B. F. Neves}
\affiliation{Departamento de F\'{\i}sica, Universidade Federal do Maranh\~{a}o,
65080-805, S\~{a}o Lu\'{\i}s, Maranh\~{a}o, Brazil.}

\begin{abstract}
We have studied BPS vortices in a CPT-odd and Lorentz-violating
Maxwell-Chern-Simons-Higgs (MCSH) electrodynamics attained from the
dimensional reduction of the Carroll-Field-Jackiw-Higgs model. The
Lorentz-violating parameter induces a pronounced behavior at origin (for the
magnetic/electric fields and energy density) which is absent in the MCSH
vortices. For some combination of the Lorentz-violating coefficients there
always exist a sufficiently large winding number $n_{0}$ such that for all $%
|n|\geq |n_{0}|$ the magnetic field flips its signal, yielding two well
defined regions with opposite magnetic flux. However, the total magnetic
flux remains quantized and proportional to the winding number.
\end{abstract}

\maketitle

\section{Introduction}

Vortex configurations constitute an important branch of research, common to
condensed matter and high energy physics. This interconnection was
established since the seminal works by Abrikosov \cite{ANO1} and
Nielsen-Olesen \cite{ANO2}, which demonstrated the existence of electrically
neutral vortices in type-II superconducting systems and in field theory
models, respectively.\ Since then, vortex solutions have become a
theoretical field of increasing interest, reinforced with the works arguing
the existence of BPS (Bogomol'nyi, Prasad, Sommerfeld) solutions \cite{BPS}.
BPS vortices were found in the Chern-Simons-Higgs model \cite{CSV} and in
the Maxwell-Chern-Simons-Higgs (MCSH)\ model \cite{MCS}, with additional
investigations involving nonminimal coupling \cite{CSV1} and other aspects
\cite{CSV2,Bolog}. Recently it has been shown the existence of generalized
Maxwell-Higgs and Chern-Simons vortices \cite{Hora1} in the context of
k-field theories \cite{kfield,GC} which have been a fertile environment for
studying new topological defects solutions \cite{kfield2}. The existence of
charged BPS vortices in a generalized Maxwell-Chern-Simons-Higgs model was
also demonstrated in Ref. \cite{Cadu-last}, while the duality between
vortices and planar Skyrmions in BPS theories has also been addressed \cite%
{Adam}. Unusual vortex configurations in condensed matter systems, endowed
with magnetic flux reversion \cite{Brevers} and fractional quantization \cite%
{Frac}, have also caught attention in the latest years.

Lorentz violating theories have been focus of strong interest since the
proposal of the standard model extension (SME)\ \cite{Colladay,Samuel},
whose gauge sector was intensively scrutinized in many respects \cite%
{KM,Klink,NM}. The study of topological defects in Lorentz-violating
scenarios was initially conducted for scalar systems defects \cite{Defects}.
The existence of monopole solutions in the presence of Lorentz violation was
regarded in the context of the Carroll-Field-Jackiw electrodynamics \cite%
{Monopole1}. Topological defects were also examined in a broader framework
of field theories endowed with tensor fields that spontaneously break the
Lorentz symmetry \cite{Seifert}.

The pioneering investigation about BPS vortex solutions in the presence of
CPT-even Lorentz-violating terms of the SME was performed in Refs. \cite%
{Carlisson1,Carlisson2}, following the idea of finding new defect solutions
in modified theoretical frameworks. In Ref. \cite{Carlisson1}, uncharged BPS
vortices were found in an Abelian Maxwell-Higgs model supplemented with
CPT-even Lorentz-violating terms belonging to the Higgs and gauge sectors of
the SME. The Lorentz-violating BPS vortices are compactlike and could
present fractional quantization of the magnetic flux. In a similar context,
it was shown that the parity-odd sector of the CPT-even term $\left(
K_{F}\right) ^{\mu \alpha \nu \beta }$ allows the existence of electrically
charged BPS vortices in absence of the Chern-Simons term \cite{Carlisson2},
endowed with magnetic flux reversion. The study of vortex configurations in
Lorentz-violating models has been an issue of active investigation recently
\cite{Belich}.

Sor far, no vortex investigation was performed in a CPT-odd and
Lorentz-violating environment. The aim of this paper is to study BPS
vortices in the Lorentz-violating planar Maxwell-Carroll-Field-Jackiw
electrodynamics, trying to answer what are the BPS vortex solutions
supported by this planar version of CPT-odd gauge sector of the SME. We
highlight the features of the corresponding BPS solutions, which possess
magnetic flux reversion: a remarkable feature induced by Lorentz-violation
which may find applications in some vortex systems of condensed matter
physics \cite{Brevers,Frac}.\ We thus start from a planar Lorentz-violating
Maxwell-Chern-Simons model attained via the dimensional reduction of the
CPT-odd Maxwell-Carroll-Field-Jackiw electrodynamics coupled to the Higgs
sector, examined in Ref. \cite{EPJC1}. In Sec. \ref{sec1}, we present the
model and implement the BPS formalism to attain the self-dual first order
equations describing the topological vortices. In Sec. \ref{sec2}, we use
the vortex ansatz to show that our solutions satisfy the usual boundary
conditions and behave as Abrikosov-Nielsen-Olesen vortices. Also, the
results of the numerical analysis are presented, revealing charged vortex
profiles that recover the MCSH solutions in the asymptotic region, and may
strongly differ from these ones near the origin. Finally, in Sec. \ref{sec3}%
, we give our remarks and conclusions.

\section{The theoretical model and BPS formalism\label{sec1}}

The Maxwell-Carroll-Field-Jackiw-Higgs model \cite{EPJC1} in $(1+3)-$%
dimensions is given by%
\begin{eqnarray}
\mathcal{L} & =&-\frac{1}{4}F_{\hat{\mu}\hat{\nu}}F^{\hat{\mu}\hat{\nu}}-%
\frac{1}{4}\epsilon^{\hat{\mu}\hat{\nu}\hat{\rho}\hat{\sigma}}\left(
k_{AF}\right) _{\hat{\mu}}A_{\hat{\nu}}F_{\hat{\rho}\hat{\sigma}}  \notag \\%
[-0.15cm]
& &  \label{L1} \\[-0.15cm]
&& +\left\vert D_{\hat{\mu}}\phi\right\vert ^{2}-V\left( \left\vert
\phi\right\vert \right) ,  \notag
\end{eqnarray}
where $\hat{\mu},\hat{\nu},\hat{\rho},\hat{\sigma}=0,1,2,3$ and $D_{\hat{\mu}%
}\phi=\partial_{\hat{\mu}}\phi-ieA_{\hat{\mu}}\phi$, the covariant
derivative. The dimensional reduction of this Lagrangian provides a kind of
MCSH model modified by Lorentz-violating (LV) terms which play the role of
coupling constants between the neutral scalar field $\left( \psi\right) $
and the abelian gauge field $(A_{\mu})$. Thus, the Lorentz-violating MCSH
model is described by the following Lagrangian density
\begin{eqnarray}
\mathcal{L} & =&-\frac{1}{4}F_{\mu\nu}F^{\mu\nu}+\frac{1}{4}s\,\epsilon
^{\nu\rho\sigma}A_{\nu}F_{\rho\sigma}+\left\vert D_{\mu}\phi\right\vert ^{2}
\notag \\
& &+\frac{1}{2}\partial_{\mu}\psi\partial^{\mu}\psi-e^{2}\psi^{2}\left\vert
\phi\right\vert ^{2}-U\left( \left\vert \phi\right\vert ,\psi\right)
\label{L2} \\
& &-\frac{1}{2}\epsilon^{\mu\rho\sigma}\left( k_{AF}\right) _{\mu}A_{\rho
}\partial_{\sigma}\psi-\frac{1}{2}\epsilon^{\mu\rho\sigma}\left(
k_{AF}\right) _{\mu}\psi~\partial_{\rho}A_{\sigma},  \notag
\end{eqnarray}
where $\psi$ is the scalar neutral field stemming from the dimensional
reduction ($\psi=A_{\hat{3}}$), the Lorentz-violating parameter $s=\left(
k_{AF}\right) _{\hat{3}}$ plays the role of a Chern-Simons coupling, $\left(
k_{AF}\right) _{\mu}$ is the (1+2)-dimensional CFJ\ vector background which\
couples\ the neutral and gauge fields. Here,\ $D_{\mu}\phi
=\partial_{\mu}\phi-ieA_{\mu}\phi$ defines the covariant derivative, while
the potential, $U\left( \left\vert \phi\right\vert ,\psi\right) $,
conveniently defined to provide BPS solutions, is
\begin{equation}
U\left( \left\vert \phi\right\vert ,\psi\right) =\frac{1}{2}\left[
ev^{2}-e\left\vert \phi\right\vert ^{2}-s\psi\right] ^{2}.
\end{equation}

The stationary Gauss's law and Ampere's law are
\begin{eqnarray}
\partial_{j}\partial_{j}A_{0}-sB-\epsilon_{ij}\left( k_{AF}\right)
_{i}\partial_{j}\psi=2e^{2}A_{0}\left\vert \phi\right\vert ^{2},
\label{Gauss_1} \\[-0.6cm]
\notag
\end{eqnarray}
\begin{eqnarray}
\epsilon_{kj}\partial_{j}B-s\epsilon_{kj}\partial_{j}A_{0}+\left(
k_{AF}\right) _{0}\epsilon_{kj}\partial_{j}\psi=eJ_{k},  \label{Ampere_1}
\end{eqnarray}
where $J_{k}$\ is the spatial component of the current density, $J^{\mu }=i%
\left[ \phi\left( D^{\mu}\phi\right) ^{\ast}-\phi^{\ast}D^{\mu}\phi\right] $.

The stationary equations of motion of the Higgs and neutral fields read
\begin{eqnarray}
0 & =& \partial_{k}\partial_{k}\phi-2ieA_{k}\partial_{k}\phi+e^{2}\left(
A_{0}\right) ^{2}\phi  \notag \\[-0.15cm]
&&  \label{Higgs_1} \\[-0.15cm]
&& -e^{2}\left( A_{k}\right) ^{2}\phi-e^{2}\psi^{2}\phi-\frac{\partial U}{%
\partial\phi^{\ast}},  \notag \\[-0.6cm]
\notag
\end{eqnarray}
\begin{eqnarray}
0 & =& \partial_{j}\partial_{j}\psi-\left( k_{AF}\right)
_{0}B-\epsilon_{ij}\left( k_{AF}\right) _{i}\partial_{j}A_{0}  \notag \\%
[-0.15cm]
& &  \label{Neutral_1} \\[-0.15cm]
& &-2e^{2}\psi\left\vert \phi\right\vert ^{2}-\frac{\partial U}{\partial\psi
},  \notag
\end{eqnarray}
respectively.

The stationary energy density $\left( \mathcal{E}\right) $\ associated with
Lagrangian (\ref{L2}), is
\begin{eqnarray}
\mathcal{E} & =&\frac{1}{2}B^{2}+\frac{1}{2}\left[ ev^{2}-e\left\vert
\phi\right\vert ^{2}-s\psi\right] ^{2}+\left\vert D_{j}\phi\right\vert ^{2}
\notag \\[-0.15cm]
& &  \label{energy_0} \\[-0.05cm]
& &\hspace{-0.5cm} +\frac{1}{2}\left( \partial_{j}A_{0}\right) ^{2}+\frac {1%
}{2}\left( \partial_{j}\psi\right) ^{2}+e^{2}A_{0}^{2}\left\vert
\phi\right\vert ^{2}+e^{2}\psi^{2}\left\vert \phi\right\vert ^{2},  \notag
\end{eqnarray}
where the condition $\left( k_{AF}\right) _{0}=0$ was imposed in order to
assuring the positiveness of the energy density. Thus, in the present
approach, only spatial components of the (1+2)-dimensional
Carroll-Field-Jackiw vector background contribute.

In the following we focus our attention on the development of a BPS
framework \cite{BPS} which provides first order differential equations
consistent with the second order equations (\ref{Gauss_1})--(\ref{Neutral_1}%
).\textbf{\ }With this aim,\textbf{\ }we first impose the following
condition on the neutral field $\psi$:
\begin{equation}
\psi=\mp A_{0},  \label{bpsCC-1}
\end{equation}
which is similar to that appearing in the context of the \textbf{\ }MCSH
vortex configurations \cite{MCS,CSV2,Bolog}. By substituting (\ref{bpsCC-1})
in Eq. (\ref{energy_0}), we achieve%
\begin{eqnarray}
\mathcal{E} & =& \frac{1}{2}B^{2}+\frac{1}{2}\left[ ev^{2}-e\left\vert
\phi\right\vert ^{2}\pm sA_{0}\right] ^{2}  \notag \\[-0.2cm]
&& \\
& &+\left\vert D_{j}\phi\right\vert ^{2}+\left( \partial_{j}A_{0}\right)
^{2}+2e^{2}A_{0}^{2}\left\vert \phi\right\vert ^{2}.  \notag
\end{eqnarray}
After converting the two first terms in quadratic form, and by using the
identity%
\begin{equation}
\left\vert D_{j}\phi\right\vert ^{2}=\left\vert D_{\pm}\phi\right\vert
^{2}\pm e\left\vert \phi\right\vert ^{2}B\pm\frac{1}{2}\epsilon_{ab}%
\partial_{a}J_{b},
\end{equation}
with $D_{\pm}\phi=D_{1}\phi\pm iD_{2}\phi$, the energy density becomes%
\begin{eqnarray}
\mathcal{E} & = &\frac{1}{2}\left[ B\mp\left( ev^{2}-e\left\vert
\phi\right\vert ^{2}\pm sA_{0}\right) \right] ^{2}  \notag \\
& &+\left\vert D_{\pm}\phi\right\vert ^{2}\pm ev^{2}B\pm\frac{1}{2}%
\epsilon_{ab}\partial_{a}J_{b} \\[0.2cm]
& &+\left( \partial_{j}A_{0}\right) ^{2}+sBA_{0}+2e^{2}A_{0}^{2}\left\vert
\phi\right\vert ^{2}.  \notag
\end{eqnarray}

By substituting Eq. (\ref{bpsCC-1}) in the Gauss law (\ref{Gauss_1}), it is
possible to transform the last three terms in a total derivative, such that
the energy density can be written as
\begin{eqnarray}
\mathcal{E} & =& \frac{1}{2}\left[ B\mp\left( ev^{2}-e\left\vert
\phi\right\vert ^{2}\pm sA_{0}\right) \right] ^{2}  \notag \\[-0.2cm]
& &  \label{E_den} \\
&& +\left\vert D_{\pm}\phi\right\vert ^{2}\pm ev^{2}B+\partial_{a}\mathcal{J}%
_{a},  \notag
\end{eqnarray}
with $\mathcal{J}_{a}$\ defined as
\begin{equation}
\mathcal{J}_{a}=\pm\frac{1}{2}\epsilon_{ab}J_{b}+A_{0}\partial_{a}A_{0}\mp%
\frac{1}{2}\epsilon_{ab}\left( k_{AF}\right) _{b}A_{0}^{2}.
\end{equation}
The energy density (\ref{E_den}) is minimized by imposing that the quadratic
terms must be null, which leads to the BPS conditions of this model,%
\begin{eqnarray}
D_{\pm}\phi=0~,  \label{BPS_cart1} \\[-0.6cm]
\notag
\end{eqnarray}
\begin{eqnarray}
B=\pm\left( ev^{2}-e\left\vert \phi\right\vert ^{2}\right) +sA_{0}.
\label{BPS_cart2}
\end{eqnarray}
which are the same ones of the MCSH model \cite{MCS}.

These BPS equations, and the Gauss law, now written as%
\begin{equation}
{\partial_{j}\partial_{j}A_{0}-sB\pm\epsilon_{ij}\left( k_{AF}\right)
_{i}\partial_{j}A}_{0}{=2e^{2}A_{0}\left\vert \phi\right\vert ^{2},}
\label{Gauss_2}
\end{equation}
describe topological vortices in this Lorentz-violating MCSH framework.

Under BPS conditions, the energy density becomes
\begin{equation}
\mathcal{E}_{BPS}=\pm ev^{2}B+\partial_{a}\mathcal{J}_{a}.  \label{den_1}
\end{equation}
After integrating under appropriated boundary conditions (see Eqs. (\ref%
{bc00}) and (\ref{bc01})), one achieves the BPS energy \
\begin{equation}
E_{BPS}={\pm}ev^{2}\int d^{2}rB={\pm}2\pi v^{2}n,  \label{den_2}
\end{equation}
which is proportional to the quantized magnetic flux, $\Phi_{B}=2\pi n/e,\ $%
where $n$\ is the winding number of the vortex configuration.

\section{Charged vortex configurations\label{sec2}}

Specifically, we look for radially symmetric solutions using the standard
static vortex Ansatz%
\begin{equation}
\phi =vg\left( r\right) e^{in\theta },~A_{\theta }=-\frac{a\left( r\right) -n%
}{er},~A_{0}=A_{0}(r),  \label{ansatz}
\end{equation}%
where $n$\ represents the winding number of the topological vortex, the
scalar functions $a\left( r\right) ,$\ $g\left( r\right) $\ and $\omega
\left( r\right) $\ are regular in $r=0$\ and at $r\rightarrow \infty $. As
usual, the fields $g\ $and $a$\ satisfy the following boundary conditions
\begin{equation}
g\left( 0\right) =0,\;a\left( 0\right) =n,  \label{bc00} \\[-0.2cm]
\end{equation}%
\begin{equation}
g\left( \infty \right) =1,\,a\left( \infty \right) =0.  \label{bc01}
\end{equation}%
The boundary conditions satisfied by the field $A_{0}$ will be explicitly
established in subsection \ref{BBCC}.

The Ansatz (\ref{ansatz}) allows to express the magnetic field in a simple
way
\begin{equation}
B=-\frac{a^{\prime }}{er}.  \label{cb1}
\end{equation}%
The BPS equations (\ref{BPS_cart1},\ref{BPS_cart2}) are rewritten as
\begin{eqnarray}
g^{\prime }& =&\pm \frac{ag}{r}\,,  \label{bq1} \\[-0.5cm]
& &  \notag
\end{eqnarray}%
\begin{equation}
B=-\frac{a^{\prime }}{er}=\pm ev^{2}\left( 1-g^{2}\right) +sA_{0}\,,
\label{bq2}
\end{equation}%
whereas the Gauss law (\ref{Gauss_2}) reads as
\begin{equation}
\hspace{-0.15cm}A_{0}^{\prime \prime }+\frac{A_{0}^{\prime }}{r}-sB\mp
\left( k_{AF}\right) _{\theta }\left( A_{0}^{\prime }+\frac{A_{0}}{2r}%
\right) -2e^{2}v^{2}g^{2}A_{0}=0,  \label{bq3}
\end{equation}%
where the upper(lower) signal corresponds to $n>0$($n<0$). \ Note that $%
\left( k_{AF}\right) _{\theta }$\ is the only component of the $(1+2)-$%
dimensional Carroll-Field-Jackiw background compatible with radially
symmetric solutions, that is, $\left( k_{AF}\right) _{\theta }$\ is the only
CFJ component remaining in the equations above describing the BPS solutions
after the Ansatz (\ref{ansatz}) is implemented. \ Note that it does not mean
the radial component $\left( k_{AF}\right) _{r}$\ was taken as null; it
simply does not participate in the formation of radially symmetric vortices.
This fact was also observed in the parity-odd coefficients $\kappa _{0i}$\
of the\ symmetric tensor $\kappa _{\mu \nu }$\ of the LV and CPT-even model\
analyzed in Ref. \cite{Carlisson2}, where only the component $\kappa
_{0\theta }$\ has\ contributed after the ansatz implementation.

From the set of equations (\ref{bq1})-(\ref{bq3}), we can observe that, for
fixed $s$\ and considering the solutions for $n>0$, the correspondent
solutions for $n<0$\ can be attained by doing $g\rightarrow g~,\
a\rightarrow -a,~A_{0}\rightarrow -A_{0}$\ and $\left( k_{AF}\right)
_{\theta }\rightarrow -\left( k_{AF}\right) _{\theta }$. We can also note
that, for $n$\ and $\left( k_{AF}\right) _{\theta }$\ fixed, under the
change $s\rightarrow -s$, the new solutions can be obtained by doing $%
g\rightarrow g~,\ a\rightarrow a~,~A_{0}\rightarrow -A_{0}$. And, by setting
$\left( k_{AF}\right) _{\theta }=0$, we recover the solutions of the MCSH
model with $s$ playing the role of the Chern-Simons parameter\textbf{.}

Using the BPS equations (\ref{bq1})-(\ref{bq2}) and\ the Gauss law (\ref{bq3}%
), we rewrite the BPS energy density (\ref{den_1}) as a sum of quadratic
terms
\begin{equation}
\mathcal{E}_{BPS}=B^{2}+2v^{2}\left( g^{\prime }\right) ^{2}+{2}%
e^{2}v^{2}\left( {g}A_{0}\right) ^{2}+\left( A_{0}^{\prime }\right) ^{2},
\label{ebps1}
\end{equation}%
showing that it is a positive-definite quantity for all values of $s$ and ${%
\left( k_{AF}\right) _{\theta }}$.

\subsection{Analysis of the boundary conditions\label{BBCC}}

We begin discussing the behavior of the solutions of Eqs. (\ref{bq1}-\ref%
{bq3}) when $r\rightarrow 0$. Using the power series method, one achieves
\begin{eqnarray}  \label{bc0_a}
g\left( r\right) & =&G_{n}r^{n}-\frac{G_{n}\left( e^{2}v^{2}+e{{\omega _{{0}}%
}}s\right) }{4}r^{n+2}  \label{bc0_g} \\[0.2cm]
&& \mp \frac{G_{n}e\omega _{0}s{\left( k_{AF}\right) _{\theta }}}{18}r^{n+3}+%
{\mathcal{\ldots }}  \notag \\[0.3cm]
a\left( r\right) & =&n-\frac{\left( e^{2}v^{2}+e{{\omega _{{0}}}}s\right) }{2%
}r^{2}\mp \frac{e\omega _{0}s{\left( k_{AF}\right) _{\theta }}}{6}r^{3}+{%
\mathcal{\ldots }}  \notag \\
\\[-0.5cm]
\notag
\end{eqnarray}
\begin{eqnarray}
A_{0}\left( r\right) & =&\omega _{0}\pm \frac{{\omega _{{0}}\left(
k_{AF}\right) _{\theta }}}{2}r  \label{bc0_w} \\[0.2cm]
&& +\frac{4s\left( ev^{2}+{{\omega _{{0}}}}s\right) +3\left[ {\left(
k_{AF}\right) _{\theta }}\right] ^{2}\omega _{0}}{16}r^{2}+{\mathcal{\ldots }%
}  \notag
\end{eqnarray}%
where $\omega _{0}=A_{0}\left( 0\right) $, and, as it happens in the usual
MCSH vortex configurations, $\omega _{{0}}$\ depends on the boundary
conditions above and it is numerically determined. The first two equations
confirm the boundary conditions imposed in (\ref{bc00}),\ while the last one
allows to impose the following condition on the field $A_{0}$\ at origin%
\begin{equation}
A_{0}^{\prime }\left( 0\right) =\pm \frac{{\omega _{{0}}\left( k_{AF}\right)
_{\theta }}}{2},  \label{bcw1}
\end{equation}

The asymptotic behavior when $r\rightarrow +\infty $ for the fields $g\left(
r\right) ,~a\left( r\right) $ and $A_{0}\left( r\right) $ is
\begin{eqnarray}
1-g\left( r\right) &\sim & r^{-1/2}e^{-\beta r},  \notag \\[0.2cm]
a\left( r\right) &\sim & r^{1/2}e^{-\beta r},  \label{Inf_1} \\[0.2cm]
A_{0}\left( r\right) &\sim & r^{-1/2}e^{-\beta r},  \notag
\end{eqnarray}%
where $\beta $ is a positive real number given by
\begin{eqnarray}
\beta _{\pm }& =&\frac{1}{2}\sqrt{\left( \kappa _{\pm }\right)
^{2}+8e^{2}v^{2}\ }-\frac{\kappa _{\pm }}{2},~\ \   \label{beta} \\[0.2cm]
\kappa _{\pm }& =&\sqrt{s^{2}+\frac{\left[ {\left( k_{AF}\right) _{\theta }}%
\right] ^{2}}{4}}\pm \frac{{\left( k_{AF}\right) _{\theta }}}{2}.
\end{eqnarray}%
where the signal $+\left( -\right) $\ stands for $n>0$\ $\left( n<0\right) $%
. It confirms the boundary conditions (\ref{bc01}) for the fields $g$, $a,$
but it also provides the boundary condition at $r\rightarrow \infty $\ for
the field $A_{0}$:%
\begin{equation}
A_{0}\left( \infty \right) =0.
\end{equation}

We observe that for fixed $s$ and ${n>0}$, the $\beta_{+}\ $parameter takes
higher values if ${\left( k_{AF}\right) _{\theta}}<0.$ We can consider two
limiting values,
\begin{equation}
\lim_{{\left( k_{AF}\right) _{\theta}\rightarrow-\infty}}\beta_{+}=ev\sqrt{2}%
,~\ \ \lim_{{\left( k_{AF}\right) _{\theta}\rightarrow+\infty}}\beta_{+}=0,
\end{equation}
between which the parameter $\beta_{+}$ varies continually. It allows to
affirm\ the profiles with $\left( k_{AF}\right) _{\theta}<0$ converge more
quickly for their saturation values than those with $\left( k_{AF}\right)
_{\theta}>0$. For $\left( k_{AF}\right) _{\theta}=0$\ and fixed $s$, the
behavior of $\beta$ is similar\ to the one of the BPS vortices coming from
the MCSH model.

\subsection{Numerical solutions}

We now introduce the dimensionless variable $\rho =evr$\ and implement the
following changes in the fields,
\begin{eqnarray}
g\left( r\right) \rightarrow \bar{g}\left( \rho \right) ,\ a\left( r\right)
\rightarrow \bar{a} \left( {\rho }\right) ,~A_{0}\left( r\right) \rightarrow
v\bar{\omega}\left( \rho \right) ,  \notag \\[-0.6cm]
\notag
\end{eqnarray}
\begin{eqnarray}
B(r)\rightarrow ev^{2}\bar{B}\left( \rho \right) ,~~\mathcal{E}%
(r)\rightarrow v^{2}\bar{\mathcal{E}}\left( \rho \right) .  \notag
\end{eqnarray}%
and rescaling the LV coefficients
\begin{equation}
\left( k_{AF}\right) _{\theta }\rightarrow ev\lambda ,~s\rightarrow ev\bar{s}%
,
\end{equation}%
where now $\lambda $ represents the (1+2)-dimensional CFJ parameter.
Thereby, the {expressions} (\ref{bq1})-(\ref{bq3}) are written in a
dimensionless form as%
\begin{eqnarray}
\bar{g}^{\prime }=\pm \frac{\bar{a}\bar{g}}{\rho }\,,  \label{eqt2} \\%
[-0.5cm]
\notag
\end{eqnarray}
\begin{eqnarray}
\bar{B}=-\frac{\bar{a}^{\prime }}{\rho }=\pm \left( 1-g^{2}\right) +\bar{s}%
\bar{\omega}\,,  \label{eqt3} \\[-0.5cm]
\notag
\end{eqnarray}
\begin{eqnarray}
\bar{\omega}^{\prime \prime }+\frac{\bar{\omega}^{\prime }}{\rho }-\bar{s}%
\bar{B}\mp \lambda \left( \bar{\omega}^{\prime }+\frac{1}{2}\frac{\bar{\omega%
}}{\rho }\right) -2\bar{g}^{2}\bar{\omega}=0\,,  \label{eqt4}
\end{eqnarray}%
while the dimensionless version of the BPS energy density is
\begin{equation}
\bar{\mathcal{E}}_{BPS}=\bar{B}^{2}+2\frac{\bar{a}^{2}\bar{g}^{2}}{\rho ^{2}}%
+{2\bar{g}^{2}\bar{\omega}}^{2}+\left( \bar{\omega}^{\prime }\right) ^{2}.
\label{H_DEF_POS}
\end{equation}

\begin{figure}[tbp]
\begin{center}
\scalebox{1}[1]{\includegraphics[width=7.5cm]{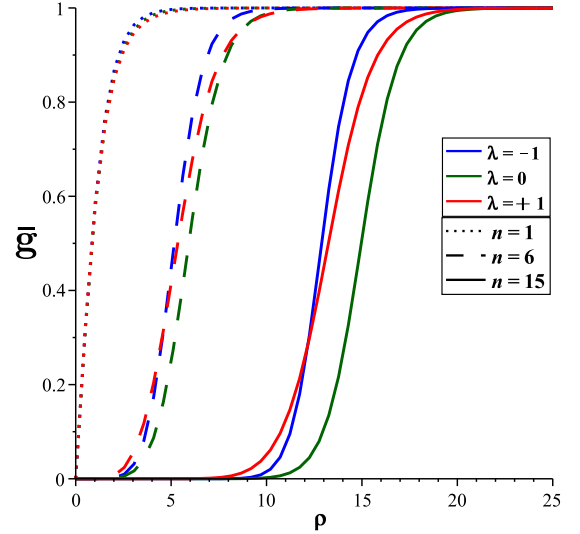}
}
\end{center}
\par
\vspace{-0.75cm}
\caption{Scalar field $\bar{g}(\protect\rho )$ (Blue lines for $\protect%
\lambda =-1$, red lines for $\protect\lambda =+1$. Green lines for $\protect%
\lambda =0$, represents the BPS solutions for the MCSH model).}
\label{S_BPS}
\end{figure}

\begin{figure}[tbp]
\begin{center}
\scalebox{1}[1]{\includegraphics[width=7.5cm]{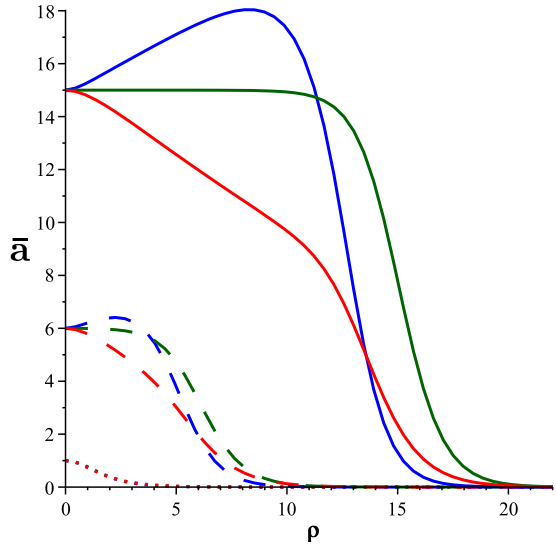}
}
\end{center}
\par
\vspace{-0.75cm}
\caption{Vector potential $\bar{a}(\protect\rho )$ (Blue lines for $\protect%
\lambda =-1$, red lines for $\protect\lambda =+1$, green lines for $\protect%
\lambda =0$).}
\label{A_BPS}
\end{figure}

\begin{figure}[tbp]
\begin{center}
\scalebox{1}[1]{\includegraphics[width=7.5cm]{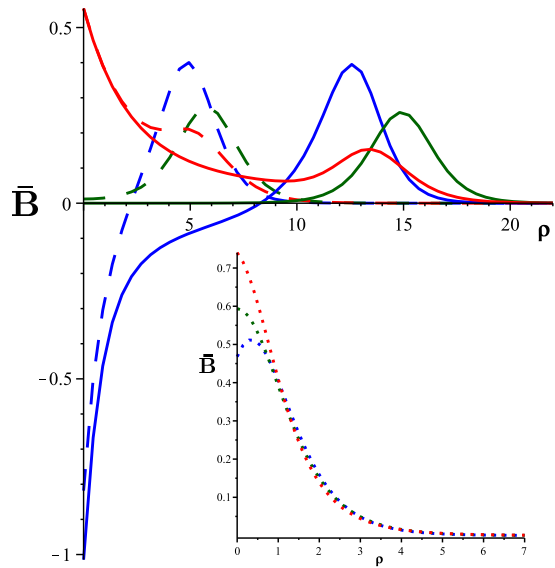}}
\end{center}
\par
\vspace{-0.75cm}
\caption{Magnetic field $\bar{B}(\protect\rho )$ (Blue lines for $\protect%
\lambda =-1$, red lines for $\protect\lambda =+1$, green lines for $\protect%
\lambda =0$).}
\label{B_BPS}
\end{figure}

\begin{figure}[tbp]
\begin{center}
\scalebox{1}[1]{\includegraphics[width=7.5cm]{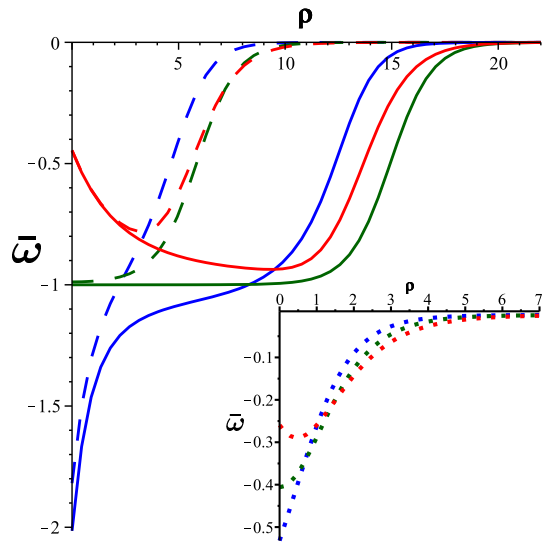}
}
\end{center}
\par
\vspace{-0.5cm}
\caption{Scalar potential $\bar{\protect\omega}(\protect\rho )$ (Blue lines
for $\protect\lambda =-1$, red lines for $\protect\lambda =+1$, green lines
for $\protect\lambda =0$. Small figure with dotted lines are the solutions
for $n=1$).}
\label{W_BPS}
\end{figure}

\begin{figure}[]
\begin{center}
\scalebox{1}[1]{\includegraphics[width=7.5cm]{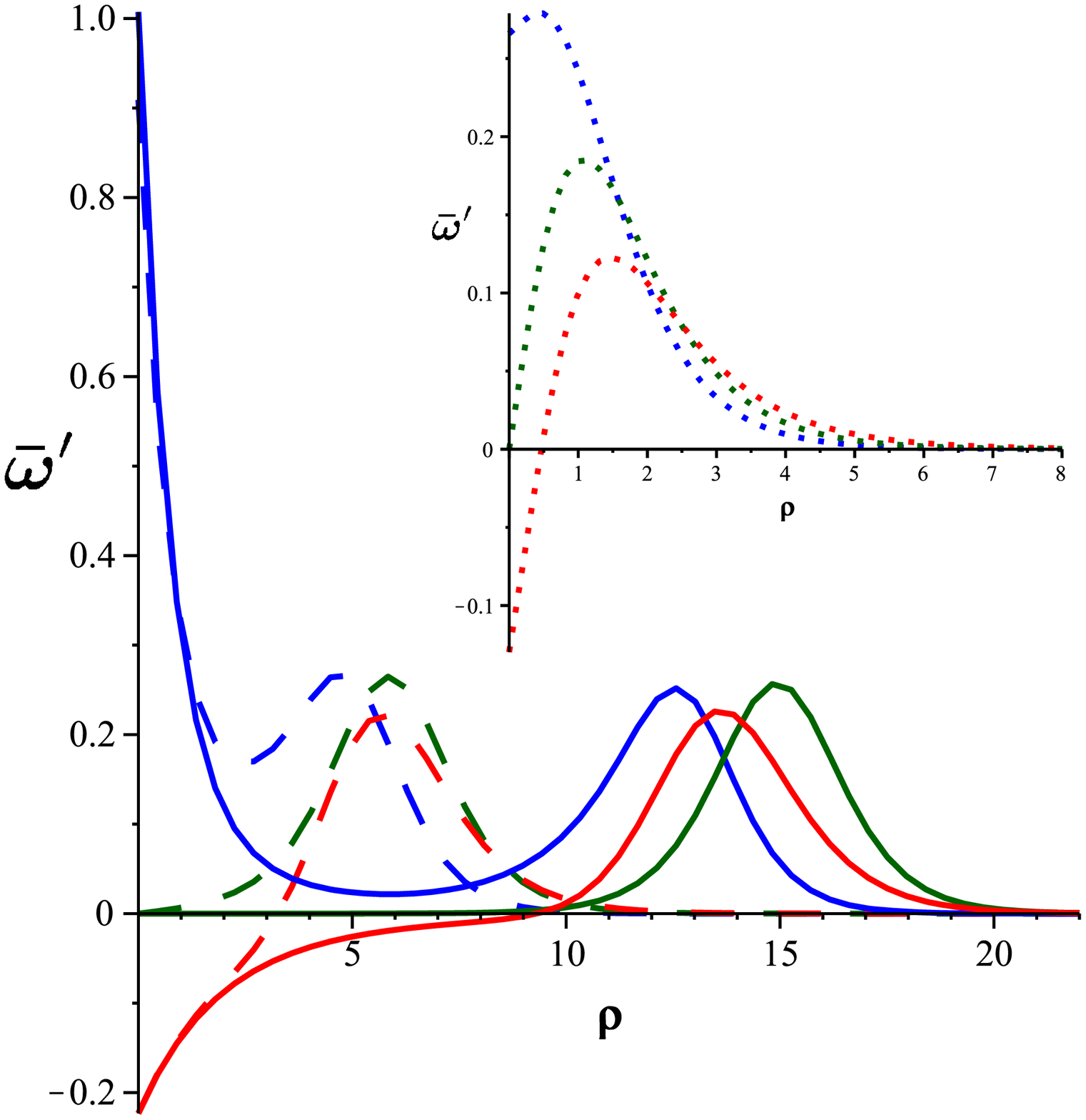}}
\end{center}
\par
\vspace{-0.5cm}
\caption{Electric field $\bar{\protect\omega}^{\prime}(\protect\rho)$. (Blue
lines for $\protect\lambda=-1$, red lines for $\protect\lambda=+1$, green
lines for $\protect\lambda=0$).}
\label{El_BPS}
\end{figure}

\begin{figure}[]
\begin{center}
\scalebox{1}[1]{\includegraphics[width=7.5cm]{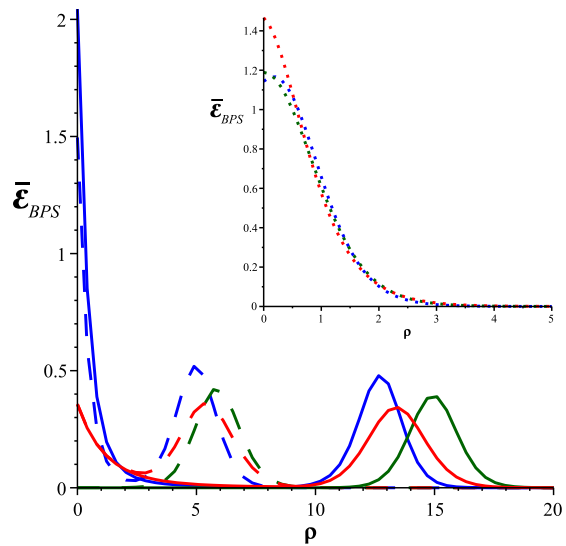}
}
\end{center}
\par
\vspace{-0.75cm}
\caption{BPS energy density $\bar{\mathcal{E}}_{_{BPS}}(\protect\rho )$.
(Blue lines for $\protect\lambda=-1$, red lines for $\protect\lambda=+1$.
Green lines for $\protect\lambda=0$).}
\label{Energy_BPS}
\end{figure}

We have performed the numerical analysis by considering three different
values for the LV parameter, $\lambda=-1,0,+1,$ and for the winding numbers,
$n=1,6,15$, while the Chern-Simons-like parameter is kept fixed, $\bar{s}=1.$
The value $\lambda=0$ reproduces the profiles of the MCSH vortices \cite%
{MCS,CSV2} which are depicted by green lines. Blue lines denote the BPS
solution with $\lambda=-1$ and red lines the ones for $\lambda=+1$. The
winding number is specified in the following way: dotted lines $\left(
n=1\right) $, dashed lines $\left( n=6\right) ,$\ and solid lines $\left(
n=15\right) $. The resultant profiles for the topological solutions are
depicted in Figs. \ref{S_BPS}--\ref{Energy_BPS}. All legends are summarized
in Fig. \ref{S_BPS}.

Fig. \ref{S_BPS} depicts the profiles of the Higgs field. For $n=1$ the
profiles are very similar to the MCSH ones ($\lambda=0$ case), but begin to
differ from them for increasing values of the winding number. In general,
the profiles with negative values of the CFJ parameter saturate more quickly
than those with positive values, in accordance with Eq. (\ref{beta}).

Fig. \ref{A_BPS} shows the profiles of the vector field.\textbf{\ }For $n=1$%
\ the profiles are very similar to the MCSH ones ($\lambda =0$\ case) but
for $n>1,$ near to the origin, the vector field magnitude can increase (for $%
\lambda =-1)$\ or decrease (for $\lambda =1)$ in relation to the MCSH\
profiles. Near to the origin, the first (second) behavior is associated a
negative (positive) magnetic field, as it is observed in Fig. \ref{B_BPS}.
For $\rho \gg 0,$\ the profiles always go to zero\ approaching the MCSH
ones. Furthermore, for $\lambda =-1$\ the vector field presents a region
with increasing magnitude, which is compatible with a sharpened negative
magnetic field around the origin. On the other hand, for $\lambda =+1$, the
vector field decreases for increasing radius, providing a positive magnetic
field which also engenders a sharpened structure around the origin.

The magnetic field behavior is represented in Fig. \ref{B_BPS}. For $n=1$
(and different $\lambda $ values) are very similar to the MCSH one, except
near the origin, where the amplitude increases with $\lambda $ \textbf{(}see
insertion with dotted lines in Fig. \ref{B_BPS}\textbf{)}. Already, for $n>1,
$\ the magnetic profiles display two localized structures that define two
well-defined domains: the first one is a pronounced magnitude flux region
centered at the origin,\ which can be positive or negative depending on the
sign of $\lambda $; the second one\ is a lump-like region whose maximum is
located in an intermediary radial distance, being compatible with a
ring-like magnetic field configuration (typical of the MCSH model) whose
radius increases with $n.$ The behavior near the origin changes strongly in
accordance with the sign of\ $\lambda $: for $\lambda =1$, $B(0)>0,$ while
for $\lambda =-1,$\ it holds $B(0)\,<0.$ Hence, for $\lambda =-1$\ there
occurs magnetic field flux inversion, once the initially negative magnetic
field\ becomes positive for $\rho >\rho _{\ast }$, with $B(\rho _{\ast })=0$%
\ (see the blue curves for $n=6$ and $n=15$\ in Fig. \ref{B_BPS}). For $%
\lambda =-1$\ \ and $n\gg 1,$\ the magnetic field saturates at the origin as
$B(0)=-1.017$.\ Further, for $n>1$\ and $\lambda =1$\ the profiles are
always positive, varying from the strong flux region centered at the origin
to the ring-like region existing for an intermediary radial coordinate. For $%
\lambda =1$ and $n\gg 1$\ the magnetic field saturates as $B(0)=0.553$.
Despite this complex scenario, the total magnetic flux is positive and
proportional to $n$, regardless the $\lambda $ value.

The scalar potential profiles, as appearing in Fig. \ref{W_BPS}, are
negative throughout the radial axis, for all values of $n$ and $\lambda $.
Near to the origin, for all $n\geq 1$, the LV profiles are different\ from
the usual MCSH solutions, with amplitudes increasing with negative $\lambda $%
.
On the other hand, far from the origin, the LV profiles closely follow the
behavior\ of the MCSH solutions (green lines).\ Near the origin, for $%
\lambda =-1,$\ the LV profiles present an inverted pronounced form whose
amplitude saturates at $\bar{\omega}(0)=-2.017$ (for $n\gg 1);$ for $\lambda
=1,$\ the profiles display a conical-shaped profile whose amplitude
saturates as $\bar{\omega}(0)=-0.446$. Note that both solutions deviate
sensibly from the MCSH $\left( \lambda =0\right) $ value at origin, $\bar{%
\omega}(0)=-1.0$. It is clear that the profile with $\lambda =-1$\ decays
faster than the one with $\lambda =1,$\ as expected.

Fig. \ref{El_BPS} contains the electric field profiles. Even for $n=1,$
these profiles are quite different at the origin neighborhood, where it
holds: $\bar{\omega}^{\prime}(0)>0$\ for $\lambda=-1$, $\bar{\omega}%
^{\prime}(0)=0$ for $\lambda=0,$\ and $\bar{\omega}^{\prime}(0)<0$\ for $%
\lambda=1$, changing a little for intermediary radius, and becoming very
similar to the MCSH ones only far from the origin (see insertion with dotted
curves in Fig. \ref{El_BPS}).
For $n>1,$ the electric field profiles display two localized significant
structures, similarly to the magnetic field behavior. In the case $%
\lambda=-1 $, the profiles are always positive, and there is a narrow cone
centered at origin, whose amplitude\ saturates as $\bar{\omega}^{\prime}(0)=$
$1.008$ $\left( \text{for }n\gg1\right) $, and a positive lump-like region
localized in an intermediary radial coordinate. On the other hand, for $%
\lambda=+1$, the electric field becomes negative near to the origin,
yielding an inverted cone with smaller amplitude, which saturates as $\bar{%
\omega }^{\prime}(0)=$ $-0.223$ for $n\gg1$. Further, as one goes away from
the origin, the electric field changes its sign, becomes positive forming a
ring-like structure around the origin, and vanishes for $\rho\rightarrow
\infty$. Note that the LV electric solutions, similarly to the magnetic
ones, differ from the usual MCSH ones mainly due the behavior near the
origin: while the MCSH solutions are null at this point, the LV parameter
may induce positive or negative electric fields in the origin neighborhood.
This pattern is shared by the BPS energy density profiles.

The BPS energy density profiles are shown in Fig. \ref{Energy_BPS}. For $n=1$%
, the profiles are lumps centered at the origin with differing amplitudes,
which overlap with each other far from $\rho =0$ (see insertion with dotted
curves in Fig. \ref{Energy_BPS}). When $n\gg 1$ and $\lambda =\pm 1$, the
BPS energy density profiles present two well-pronounced regions, in
accordance with the profiles of the magnetic and electric fields. The first
one is a peak centered at the origin whose amplitude\ saturates at $0.356$ $%
\left( \text{for }\lambda =+1\right) $ and at $2.052$ $\left( \text{for }%
\lambda =-1\right) $. This concentration of energy at the origin is not
present in the MCSH model. The second\ one is a lump-shaped region located
at an intermediary distance from the origin, whose radius increases with $n,$
as it happens in the MCSH model. For each value of $n,$ the ring radius
follow the inequality: $\rho _{\lambda =-1}\lesssim \rho _{\lambda
=1}\lesssim \rho _{\lambda =0}$.

\section{Conclusions and remarks \label{sec3}}

In this work, we have considered a Lorentz-violating planar electrodynamics,
attained from the dimensional reduction of the
Maxwell-Carroll-Field-Jackiw-Higgs model, as a theoretical environment for
studying charged BPS vortex configurations. After writing the stationary
equations of motion and the energy density, the radially symmetric usual
vortex Ansatz was implemented, keeping only one Lorentz-violating parameter
in the equations of the system. By manipulating the energy density, the BPS
equations were obtained, confirming the existence of BPS solutions for such
a model. The BPS energy remains connected with the magnetic flux
quantization, as it is usual.

The numerical simulations were performed for different values of the
Lorentz-violating parameter, and distinct winding numbers.\ For $n=1$, at
the origin, the solutions sometimes present notable deviations from the MCSH
profiles. However for large radius they closely follow the behavior of the
MCSH ones. In general, the deviation\ is more accentuated for increasing
winding number values. \ For $n>1$, the LV profiles keep some similarity to
the usual MCSH\ solutions in the large radius region but decay\ in
accordance\ with its respective mass scale:\ $\beta _{\lambda <0}>\beta
_{\lambda =0}>\beta _{\lambda >0}$. \ The role played by the LV\ parameter
becomes more pronounced at the origin, where it generates a peaked profile
(absent in the MCSH solutions) in the magnetic/electric field and the energy
density profiles. When $\lambda <0$ the magnetic field assumes negative
values near the origin. However, for a sufficiently large radius, it flips
its signal, providing two regions with opposite magnetic flux. The flipping
of the magnetic flux represents a remarkable feature induced by
Lorentz-violation,\ which may find applications in condensed matter systems
endowed with magnetic flux reversion \cite{Brevers,Frac}. This feature was
also observed in charged vortex configurations defined in the context of the
CPT-even and nonbirefringent Lorentz-violating model of Ref. \cite%
{Carlisson2}.

The analysis performed above can be extended for all values of $\bar{s}$\
and $\lambda$. For $n>0$, a fixed $\bar{s}$,$~$and $\lambda>0,$\ there are
always two well defined regions with positive magnetic flux, occurring no
magnetic field reversion. On the other hand, for\ fixed $\bar{s}$\ and $%
\lambda <0,$\ there always exists a sufficiently large winding number $n_{0}$%
\ such that for $n>n_{0}$\ the magnetic field reverses its signal.
Consequently, there are always two well defined regions with opposite
magnetic flux. A similar result is obtained for $n<0$. However, in all cases
the total magnetic flux remains quantized and proportional to the winding
number.

\textbf{Acknowledgements}

The authors thank to CAPES, CNPq and FAPEMA (Brazilian agencies) for
financial support.

\end{document}